\documentclass[conference]{IEEEtran}
\usepackage{cite}
\usepackage{graphicx}
\usepackage[cmex10]{amsmath}
\usepackage{amsfonts,amssymb,amsthm}
\usepackage{array,multirow}
\usepackage{float}
\usepackage{balance}
\usepackage[english]{babel}
\usepackage{xcolor}
\usepackage[normalem]{ulem} 
\usepackage{enumitem}
\usepackage[caption=false, font=footnotesize]{subfig}
\usepackage{nicefrac}

\def\BibTeX{{\rm B\kern-.05em{\sc i\kern-.025em b}\kern-.08em
    T\kern-.1667em\lower.7ex\hbox{E}\kern-.125emX}}

\begin{document}

\title{\LARGE The Impact of 5G Channel Models on the Performance of Intelligent Reflecting Surfaces and Decode-and-Forward Relaying}

\author{\IEEEauthorblockN{Ioannis~Chatzigeorgiou}
\IEEEauthorblockA{School of Computing and Communications\\
Lancaster University\\Lancaster LA1 4WA, United Kingdom\\
Email: i.chatzigeorgiou@lancaster.ac.uk}}

\maketitle

\begin{abstract}
An intelligent reflecting surface (IRS) is an array of discrete elements with configurable scattering properties. It has the capability to beamform arriving radio waves to an intended receiver, making it an attractive candidate technology for fifth-generation (5G) communications. A recent study debunked the notion that IRSs can replace relays because a large number of IRS elements is required even to approach the performance of simple single-antenna decode-and-forward (DF) relays. The study introduced 4G channel models into a theoretical framework to obtain simulation results, based on which comparisons between the two schemes were carried out. In this paper, we consider 5G channel models, reflect on the revised results, and argue that IRSs and DF relays can complement each other's strengths and can both have a place in 5G and beyond 5G architectures.
\end{abstract}

\vspace{4pt}
\begin{IEEEkeywords}
Path loss, channel model, millimeter waves, intelligent reflecting surface, decode-and-forward relaying.
\end{IEEEkeywords}

%%%%%%%%%%%%%%%%%%%
% Section 1. Introduction
%%%%%%%%%%%%%%%%%%%
\section{Introduction}
\label{sec:intro}

Intelligent reflecting surfaces (IRSs)~\cite{Wu2019}, also referred to as reconfigurable intelligent surfaces~\cite{Huang2019} and software-defined hyper-surfaces~\cite{Liaskos2018}, are considered for fifth-generation (5G) and beyond 5G architectures because of their adjustable scattering properties. An IRS is an array of sub-wavelength-sized elements, which act as diffuse scatterers that align the phases of the reflected signals and focus them toward the desired direction~\cite{Ozdogan2020}. IRS technology has thus the potential to control and optimize the wireless propagation environment between a transmitter and a receiver.

Bj\"{o}rnson~\textit{et al.}~\cite{Bjornson2020} consider an ideal IRS of multiple elements and a repetition-coded decode-and-forward (DF)~relay, which is equipped with a single antenna of size equal to that of an IRS element. They derive closed-form expressions for the achievable rates and the required transmission powers of each scheme, which help them determine the number of elements that an IRS should have in order to outperform DF relaying. Although DF relaying suffers from a pre-log penalty in the rate due to the two-hop transmission, results demonstrate that the IRS needs a very large number of elements to match the performance of a single-antenna DF relay. Consequently, the authors argue that the benefits of IRS-aided transmission do not outweigh the benefits of conventional DF relaying, and the former does not have a strong case for replacing the latter.

The radio propagation models that were used in~\cite{Bjornson2020} have been defined by the 3rd generation partnership project (3GPP) in the technical specification for the evolved universal terrestrial radio access (E-UTRA)~\cite{3GPP_TS_36_814}, which is part of the long-term evolution (LTE) and LTE-Advanced (LTE-A). Although 4G will most likely remain the dominant mobile communications technology for a few more years, many countries have already started to upgrade their network infrastructure to support 5G services. The objective of this paper is to introduce 5G channel models, also defined by 3GPP~\cite{3GPP_TR_38_901}, in the simulation setup of~\cite{Bjornson2020}, look into how the change in the channel models affects the results presented in~\cite{Bjornson2020}, and identify scenarios where the emerging IRS technology can coexist with classic relaying.

The remainder of this paper has been organized as follows: Section~\ref{sec:system} presents the system model, describes transmission aided by a DF relay or an IRS, and mentions expressions for the transmit power that is required to support a particular rate, as derived in~\cite{Bjornson2020}. Section~\ref{sec:channels} gives an overview of outdoor path loss models proposed for 5G networks. Simulation results and performance trade-offs between relay-aided and IRS-aided transmission are discussed in Section~\ref{sec:comparison}, and conclusions are drawn in Section~\ref{sec:conclusions}.

%%%%%%%%%%%%%%%%%%%
%  Section 2. System Model and Power Requirements
%%%%%%%%%%%%%%%%%%%
%\vspace{-8pt}
\section{System Model and Power Requirements}
\label{sec:system}

\begin{figure}[t]
\centering
\includegraphics[width=0.7\columnwidth]{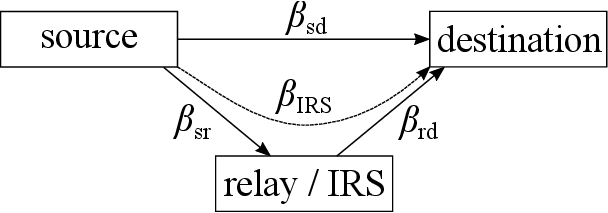}
\caption{System model comprising a source, a relay or IRS and a destination. The channel gains between the nodes are $\beta_\mathrm{sd}$, $\beta_\mathrm{sr}$ and $\beta_\mathrm{rd}$, while $\beta_\mathrm{IRS}$ is the composite channel gain when an IRS is used.}
\label{fig:system_model}
%\vspace{-9pt}
\end{figure}

We consider a system consisting of a single-antenna source and a single-antenna destination. Transmission from the source to the destination is aided either by a single-antenna repetition-coded DF relay or an IRS, as shown in Fig.~\ref{fig:system_model}.

When a half-duplex DF relay is used, the transmission is split into two stages. In the first stage, the source broadcasts a signal to the relay and the destination. In the second stage, the relay decodes, re-encodes, and transmits the signal to the destination. At the end of the two-stage process, the destination combines the two copies of the received signal using maximum ratio combining.

When an IRS is used, the transmission is completed in a single stage. The source broadcasts a signal to both the IRS and the destination, and the IRS uses $N$ elements to reflect and direct the incoming signal toward the destination. The phase shifts of the $N$ discrete elements can be optimized, so that the $N$ reflected signals and the signal transmitted by the source are constructively added at the destination~\cite[Lemma 1]{Bjornson2020}.

If $\bar{R}$ is the target rate of the system, the required transmit power at the source, when a DF relay or an IRS is deployed, can be computed using~\cite[Corollary 1]{Bjornson2020}:
\begin{equation}
p_\mathrm{DF}= \begin{cases}  
\left(2^{2\bar{R}}-1\right)\frac{\sigma^2}{\beta_\mathrm{sd}} &\text{if $\beta_\mathrm{sd} > \beta_\mathrm{sr}$,} \\
\left(2^{2\bar{R}}-1\right)\frac{(\beta_\mathrm{sr}+\beta_\mathrm{rd}-\beta_\mathrm{sd})\sigma^2}{2\beta_\mathrm{sr}\beta_\mathrm{rd}} &\text{if $\beta_\mathrm{sd} \leq \beta_\mathrm{sr}$} 
\end{cases}
\label{eq:power_DF}
\end{equation}
\begin{equation}
p_\mathrm{IRS}(N)=\left(2^{\bar{R}}-1\right)\frac{\sigma^2}{(\sqrt{\beta_\mathrm{sd}}+N\alpha\sqrt{\beta_\mathrm{IRS}})^2},
\label{eq:power_IRS}
\end{equation}
where $\sigma^2$ is the power of the additive white Gaussian noise at the destination, $\alpha\!\in\!(0,1]$ is the amplitude reflection coefficient of the IRS, and $\beta_\mathrm{sr}$, $\beta_\mathrm{rd}$, $\beta_\mathrm{sd}$ are the squared magnitudes of the deterministic fading gains of the channels between the respective nodes. In the case of the IRS, $\beta_\mathrm{IRS}$ is the squared average of the products between the magnitudes of the fading gains of the input and output channels of every IRS element. If each IRS element has the same size as the antenna of the relay, the magnitude of the fading gain of every input channel will be $\sqrt{\beta_\mathrm{sr}}$ and the magnitude of the fading gain of every output channel will be $\sqrt{\beta_\mathrm{rd}}$. We can thus write:
\begin{equation}
\beta_\mathrm{IRS}=\left(\frac{1}{N}\left(N\sqrt{\beta_\mathrm{sr}}\sqrt{\beta_\mathrm{rd}}\right)\right)^{\!2}=\beta_\mathrm{sr}\beta_\mathrm{rd}.
\end{equation}
For comparison with \eqref{eq:power_DF} and \eqref{eq:power_IRS}, the required transmit power for a target rate $\bar{R}$ without the assistance of a relay/IRS is:
\begin{equation}
p_\mathrm{SISO}=\left(2^{\bar{R}}-1\right)\frac{\sigma^2}{\beta_\mathrm{sd}}.
\end{equation}

\begin{table}[t]
\renewcommand{\arraystretch}{1.3}
\caption{Path loss models for urban microcell when $d_\mathrm{2D}$ is shorter than the breakpoint distance $d_\mathrm{BP}$. Distances are in meters, $f_\mathrm{c}$~is~in~GHz and $h_\mathrm{BS}=10\mathrm{m}$ \cite[Table 7.4.1-1]{3GPP_TR_38_901}.}%
\centering 
\begin{tabular}{|c|l|}% 
\hline 
\parbox[t]{2mm}{\multirow{3}{*}{\rotatebox[origin=c]{90}{LOS}}}
& $PL_\mathrm{LOS}=32.4+21\log_{10}(d_\mathrm{3D})+20\log_{10}(f_\mathrm{c})$,\\
& \multicolumn{1}{r|}{for $10\mathrm{m}\leq d_\mathrm{2D}\leq d_\mathrm{BP}$}\\
& \multicolumn{1}{r|}{where $d_\mathrm{BP}=36(h_\mathrm{UT}-1)f_\mathrm{c}/(3\times 10^{-1})$ for $h_\mathrm{BS}=10\mathrm{m}$}\\
\hline
\parbox[t]{2mm}{\multirow{6}{*}{\rotatebox[origin=c]{90}{NLOS}}}
& $PL_\mathrm{NLOS}=\max\bigl(PL_\mathrm{LOS}, PL'_\mathrm{NLOS}\bigr)$,\\
& where:\\
& $PL'_\mathrm{NLOS}=22.4+35.3\log_{10}(d_\mathrm{3D})+21.3\log_{10}(f_\mathrm{c})$\\
& \qquad\qquad\quad\; $-0.3(h_\mathrm{UT}-1.5)$,\\
& \multicolumn{1}{r|}{for $10\mathrm{m}\leq d_\mathrm{2D}\leq 5\mathrm{km}$}\\
& \multicolumn{1}{r|}{and $1.5\mathrm{m}\leq h_\mathrm{UT}\leq 22.5\mathrm{m}$}\\
\hline
\end{tabular}
\label{tb:UMi_models}
%\vspace{-7pt}
\end{table}

If $\beta_\mathrm{sd}\leq\beta_\mathrm{sr}$, the required transmit power in the case of the IRS-aided system is lower than that of the relay-aided system, only if $N>N_\textrm{min}$, where~\cite[Proposition 2]{Bjornson2020}:
\begin{equation}
N_\mathrm{min}=%
\frac{%
\sqrt{%
\left(\sqrt{1+\frac{2p_\mathrm{DF}\beta_\mathrm{sr}\beta_\mathrm{rd}}{(\beta_\mathrm{sr}+\beta_\mathrm{rd}-\beta_\mathrm{sd})\sigma^2}}-1
\right)%
\frac{\sigma^2}{p_\mathrm{DF}}%
}%
-\sqrt{\beta_\mathrm{sd}}
}%
{\alpha\sqrt{\beta_\mathrm{IRS}}}.
\end{equation}
Otherwise, if $\beta_\mathrm{sd}>\beta_\mathrm{sr}$, the required transmit power in the IRS case will be lower than that of the DF case, for any $N\geq 1$.

This section described the system model and presented key expressions that were derived in \cite{Bjornson2020}. The following section discusses how path loss models can be introduced in the calculation of the channel gains $\beta_\mathrm{sr}$, $\beta_\mathrm{rd}$, and $\beta_\mathrm{sd}$.

%%%%%%%%%%%%%%%%%%%
%  Section 3. Channel Models for Urban Environments
%%%%%%%%%%%%%%%%%%%
%\vspace{-8pt}
\section{Channel Models for Urban Environments}
\label{sec:channels}

\begin{figure}[t]
\centering
\includegraphics[width=0.5\columnwidth]{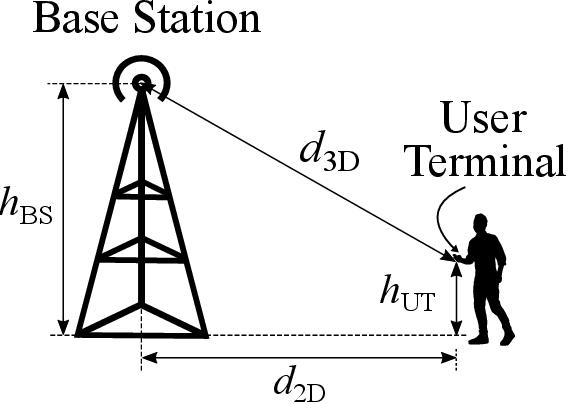}
\caption{Height and distance definitions for outdoor urban scenarios \cite{3GPP_TR_38_901}.}
\label{fig:BS_and_UT}
\end{figure}

An overview of propagation models for 5G systems operating at frequencies in the range $0.5$-$100$ GHz is given in~\cite{Rappaport2017}. The overview includes models considered in 3GPP~TR~38.901~\cite{3GPP_TR_38_901}. This section focuses on outdoor-to-outdoor radio propagation in urban environments, as shown in Fig.~\ref{fig:BS_and_UT}, and summarizes path loss models for urban canyons, described in~\cite{3GPP_TR_38_901}.

The \textit{urban microcell} (UMi) case in~\cite{3GPP_TR_38_901} captures scenarios where base stations are mounted below the rooftop levels of surrounding buildings. Radio propagation in a canyon-like environment is assumed, that is, users move along streets flanked by buildings on both sides. Typical values for the height of the base station (BS) and the height of the user terminal (UT) are $h_\mathrm{BS}=10\mathrm{m}$ and $1.5\mathrm{m}\leq h_\mathrm{UT}\leq 22.5\mathrm{m}$, respectively. Path loss models for line-of-sight (LOS) and non-line-of-sight (NLOS) propagation are presented in Table~\ref{tb:UMi_models}. For path loss calculations, the height $h_\mathrm{UT}$ and the distance $d_\mathrm{3D}$ should be expressed in meters. The carrier frequency $f_\mathrm{c}$ should be in GHz and can take values in the range of $0.5\leq f_\mathrm{c}\leq 100$.

\begin{figure}[t]
\centering
\includegraphics[width=0.67\columnwidth]{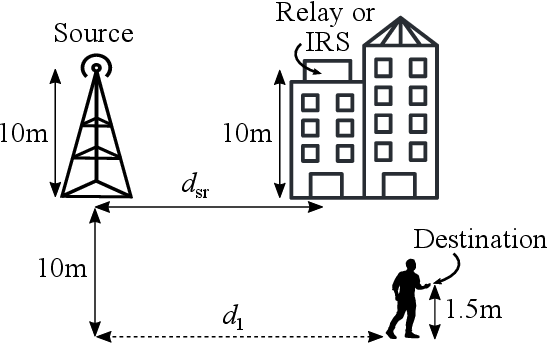}
\caption{Configuration of system simulation, where $0\,\mathrm{m}\leq d_1 \leq 160\,\mathrm{m}$.}
\label{fig:config}
%\vspace{-7pt}
\end{figure}

In order to obtain a deterministic model for the channel gain $\beta$, shadow fading has been neglected, as in~\cite{Bjornson2020}. The channel gain $\beta$ can be expressed as a function of $h_\mathrm{UT}$, $d_\mathrm{3D}$, and $f_\mathrm{c}$ as follows:
\begin{equation}
\label{eq:channel_gain_model}
\begin{split}
\beta(h_\mathrm{UT}, &d_\mathrm{3D}, f_\mathrm{c})\,\mathrm{[dB]}=\\
&=G_\mathrm{BS}\,\mathrm{[dBi]}+G_\mathrm{UT}\,\mathrm{[dBi]}+PL(h_\mathrm{UT}, d_\mathrm{3D}, f_\mathrm{c}),
\end{split}
\end{equation}
where $G_\mathrm{BS}$ and $G_\mathrm{UT}$ are the antenna gains at the base station and the user terminal, respectively. The path loss expression for $PL(h_\mathrm{UT}, d_\mathrm{3D}, f_\mathrm{c})$ can be obtained from Table~\ref{tb:UMi_models}, depending on the propagation model (LOS or NLOS).

%%%%%%%%%%%%%%%%%%%
%  Section 4. Performance Comparison and Discussion
%%%%%%%%%%%%%%%%%%%
%\vspace{-8pt}
\section{Performance Comparison and Discussion}
\label{sec:comparison}

\begin{table}[t]
\renewcommand{\arraystretch}{1.5}
\caption{Path loss models and values for the input arguments of \eqref{eq:channel_gain_model} for the calculation of $\beta_\mathrm{sr}$, $\beta_\mathrm{rd}$ and $\beta_\mathrm{sd}$. The antenna gains have been set to $8\,\mathrm{dBi}$, while $1.35\,\mathrm{GHz}\leq f_\mathrm{c}\leq 100\,\mathrm{GHz}$.}%
\centering 
\begin{tabular}{|c|c|c|c|}% 
\hline 
&Model & $h_\mathrm{UT}$  & $d_\mathrm{3D}$\\
\hline\hline
$\beta_\mathrm{sr}$ & LOS & 10m & $d_\mathrm{sr}$\\
\hline
$\beta_\mathrm{rd}$ & LOS & 1.5m & $\sqrt{(d_1-d_\mathrm{sr})^2 +172.25}$ \\
\hline
$\beta_\mathrm{sd}$  & NLOS & 1.5m & $\sqrt{(d_1)^2+172.25}$\\
\hline
\end{tabular}
\label{tb:parameters_for_experiment}
\end{table}

The simulation setup is a three-dimensional extension of the two-dimensional setup in~\cite{Bjornson2020}. The source and the relay/IRS are laid $d_\mathrm{sr}=80$m apart at fixed positions, while the destination moves along a line that is parallel to the line connecting the source and the relay/IRS. The two lines are separated by $10$m, so that the minimum value requirement for $d_\mathrm{2D}$, given in Table~\ref{tb:UMi_models}, is met. The destination covers a distance, denoted by $d_1$, which gradually increases from $0$m to $160$m. The heights of the source, relay/IRS and destination have been set to $10$m, $10$m and $1.5$m, respectively. The spatial configuration of the system is depicted in Fig.~\ref{fig:config}.

Similarly to \cite{Bjornson2020}, LOS propagation is assumed from the source to the relay/IRS and from the relay/IRS to the destination. To motivate and justify the use of a relay/IRS, the source to destination link experiences NLOS conditions. Equation~\eqref{eq:channel_gain_model} for the channel gain $\beta$ can be used for any transmitter-receiver pair, including the source-relay, relay-destination and source-destination pairs. Table~\ref{tb:parameters_for_experiment} shows how the parameters of \eqref{eq:channel_gain_model}~need to be configured to obtain $\beta_\mathrm{sr}$, $\beta_\mathrm{rd}$ and $\beta_\mathrm{sd}$, while $\beta_\mathrm{IRS}=\beta_\mathrm{sr}\beta_\mathrm{rd}$. Antenna gains have been set to $8$ dBi~\cite[Table 7.3-1]{3GPP_TR_38_901} at the source and the relay/IRS, and to $0$ dBi at the destination. The values for $f_\mathrm{c}$ have been restricted to the range $[1.35,100]$ GHz. When LOS propagation is possible, this constraint ensures that the two-dimensional distance $d_\mathrm{2D}$ between the relay/ISR and the destination will not be greater than the breakpoint distance $d_\mathrm{BP}$. The two-dimensional distance between the relay/ISR and the destination will not exceed $\sqrt{80^2+10^2}\approx 80.6\mathrm{m}$ in the simulation, which is achieved for $d_1=0\mathrm{m}$ and $d_1=160\mathrm{m}$. For $f_\mathrm{c}\geq 1.35\;\mathrm{GHz}$, we get $d_\mathrm{BP}\geq 81\;\mathrm{m}$, therefore $d_\mathrm{2D}<d_\mathrm{BP}$ for any $d_1\in[0, 160]\;\mathrm{m}$, as per the requirement in Table~\ref{tb:UMi_models}.  

\begin{figure}[!t]
\subfloat[LTE-A for the UMi case, as shown in \cite{Bjornson2020}]{%
\includegraphics[width=8.75cm]{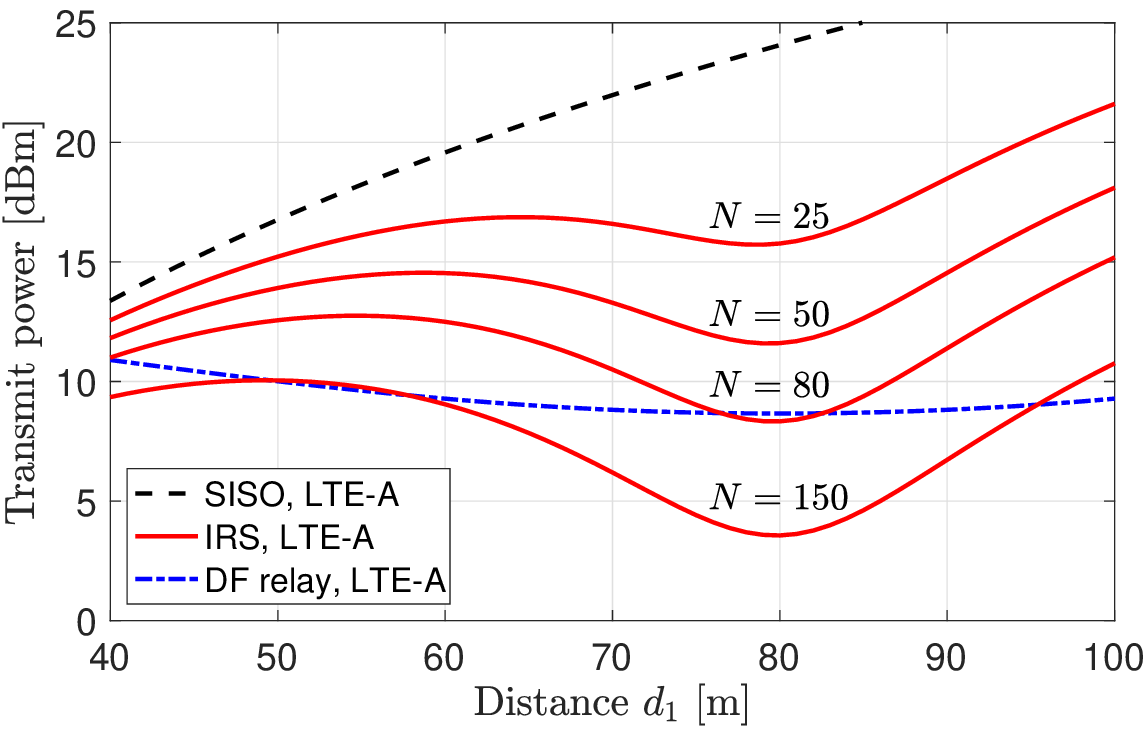}
\label{fig:LTE_A}}
\\
\subfloat[5G for the UMi case]{%
\includegraphics[width=8.75cm]{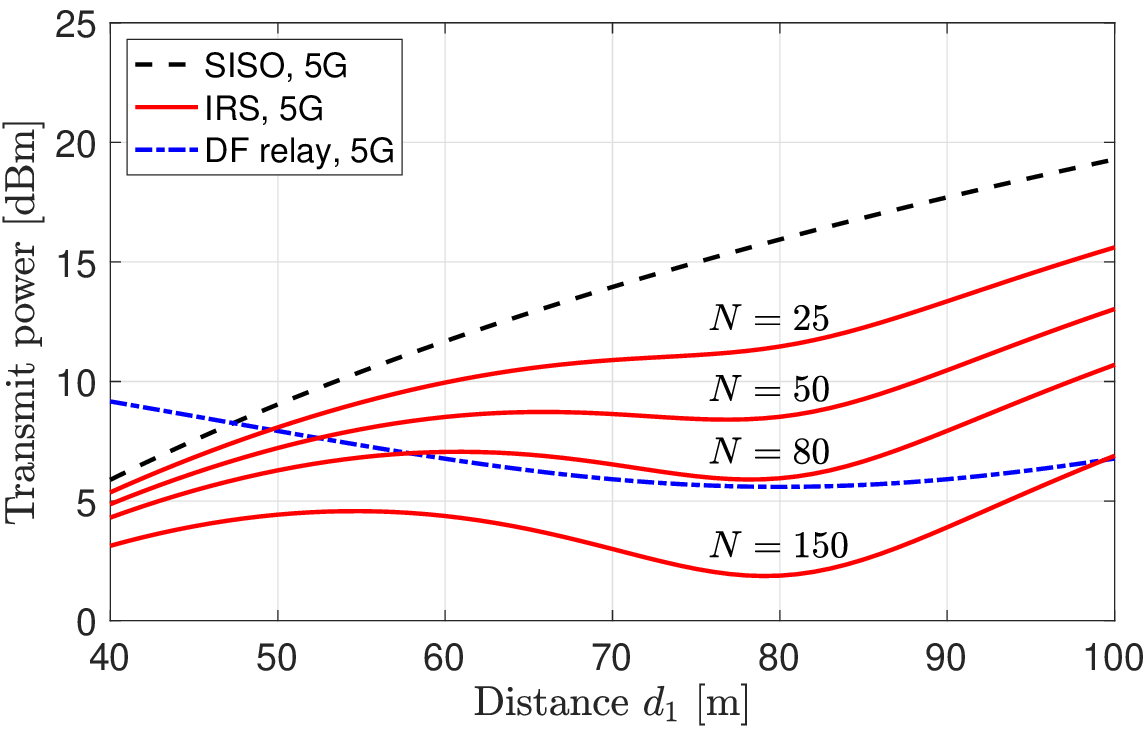}
\label{fig:5G}}%
\caption{Required transmit power to support $\bar{R}=6$ bits/sec/Hz as a function of $d_1$ for channel models specified in (a) E-UTRA (LTE-A)~\cite{3GPP_TS_36_814} and (b) 5G~\cite{3GPP_TR_38_901}. The carrier frequency has been set to $f_\mathrm{c}=3$ GHz, the bandwidth is $B=10$ MHz and $d_\mathrm{sr}=80$m.}
\label{fig:LTE_A_vs_5G}
\end{figure}

The transmit power that is needed to achieve a rate of $\bar{R}=6$ bits/sec/Hz, for channel models defined for LTE-A and 5G, is presented in Fig.~\ref{fig:LTE_A_vs_5G}. An IRS having $N=\{25,50,80,150\}$ elements has been considered for $\alpha=1$. The carrier frequency is $f_\mathrm{c}\!=3\!$ GHz, the bandwidth is $B\!=\!10\!$ MHz and the resultant noise power is $-94$ dBm. In both Fig.~\ref{fig:LTE_A} and Fig.~\ref{fig:5G}, similar trends can be observed; as $d_1$ increases, the SISO case needs the highest power, while the IRS case requires a lower transmit power than DF relaying for a broader range of $d_1$ values, only when $N$ is large. A notable difference between LTE-A and 5G for $N=80$ is that, when the destination is opposite to the relay/IRS ($d_1=d_\mathrm{sr}=80$m), the IRS case consumes less transmit power than DF relaying in LTE-A but not in 5G. However, the gap between the IRS and DF curves is smaller in 5G. Furthermore, IRS outperforms both SISO and DF relaying for $d_1<50$m, when the 5G channel models are used.

\begin{figure}[t]
\centering
\includegraphics[width=8.75cm]{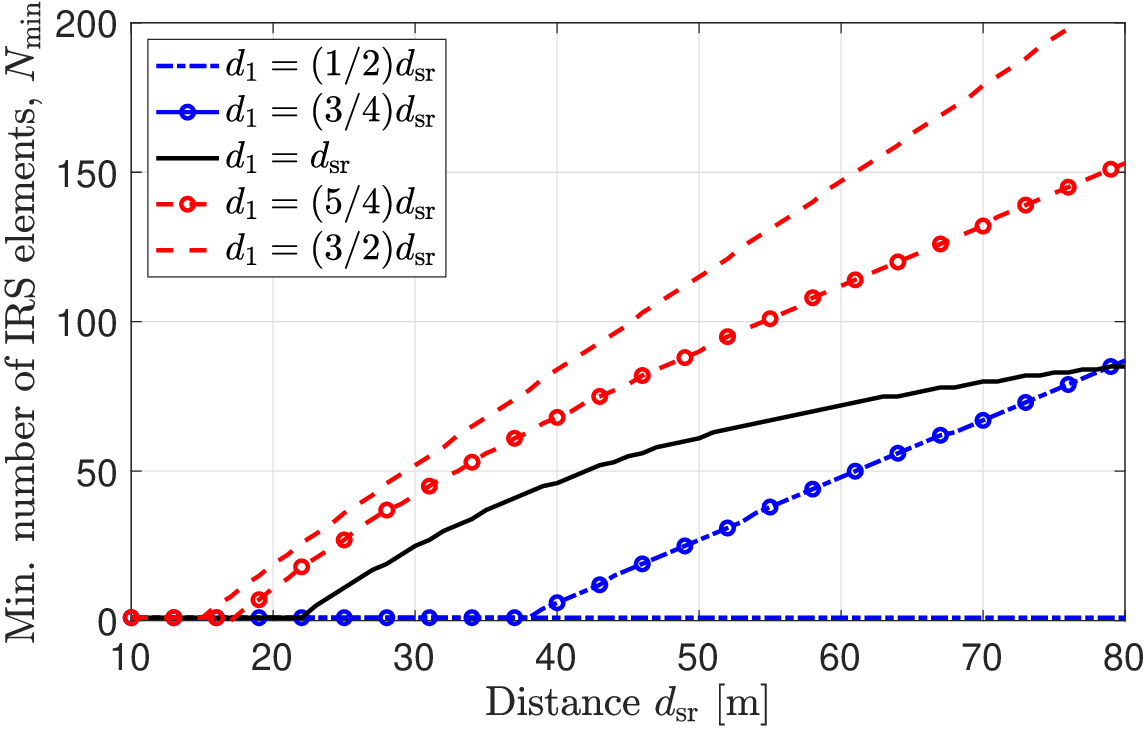}
\caption{Minimum number of IRS elements $(N_\mathrm{min})$ for IRS to support $\bar{R}=6$ bits/sec/Hz for a lower transmit power than DF relaying and SISO. Number $N_\mathrm{min}$ is depicted as a function of $d_\mathrm{sr}$ for $f_\mathrm{c}=3$ GHz, $B=10$~MHz and various values of $d_1$.}
\label{fig:dsr_vs_Nmin}
\end{figure}

The impact of $d_1$ and $d_\mathrm{sr}$ on the performance of ISR is further investigated in Fig.~\ref{fig:dsr_vs_Nmin}, where $d_\mathrm{sr}$ ranges from $10$m to $80$m and $d_1=\{(\nicefrac{1}{2})d_\mathrm{sr},(\nicefrac{3}{4})d_\mathrm{sr},(\nicefrac{5}{4})d_\mathrm{sr},(\nicefrac{3}{2})d_\mathrm{sr}\}$. The figure shows the minimum required number of elements for the IRS case to achieve a transmit power that is lower than that of SISO and DF relaying when $\bar{R}=6$ bits/sec/Hz. We observe that, for $d_1=d_\mathrm{sr}/2$, the IRS case outperforms SISO and DF relaying across the $d_\mathrm{sr}$ range for any $N\geq 1$. As the ISR moves away from the source and $d_\mathrm{sr}$ increases, more elements are needed as $d_1$ approaches and exceeds the value of $d_\mathrm{sr}$. Fig.~\ref{fig:5G} and Fig.~\ref{fig:dsr_vs_Nmin} establish that a large number of elements is required to extend the coverage of the source, i.e., when $d_1>d_\mathrm{sr}$, making DF relaying the best option. However, IRS is a viable solution when the objective is to reduce the transmit power at the source for a given coverage, i.e., when $d_1\leq d_\mathrm{sr}$, provided that $d_\mathrm{sr}$ is smaller than a value that depends on $N$, $f_\mathrm{c}$ and $\bar{R}$.

\begin{figure}[t]
\centering
\includegraphics[width=8.75cm]{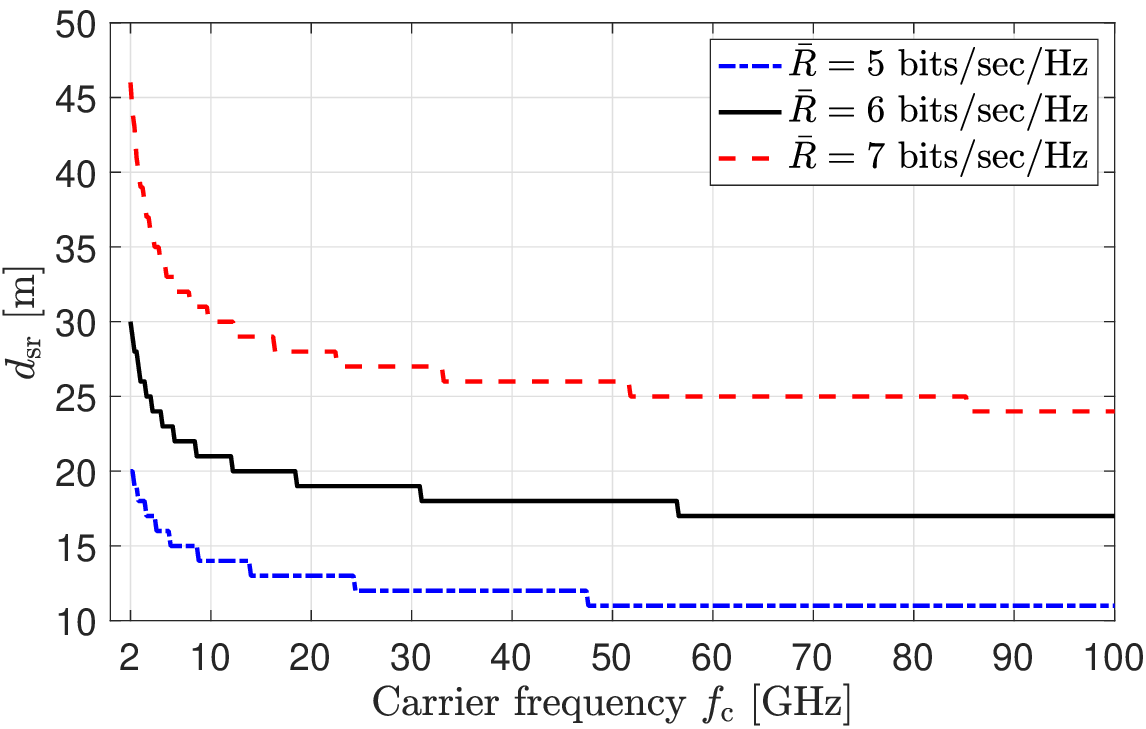}
\caption{Maximum distance that an IRS of $N=16$ elements can be placed away from the source, so that a lower transmit power than that of DF relaying and SISO can be used to offer a rate $\bar{R}$ of $5$, $6$ or $7$ bits/sec/Hz to a destination located at $d_1\in[d_\mathrm{sr}/2,d_\mathrm{sr}]$. The bandwidth is $B=10$ MHz.
}
\label{fig:dsr_vs_fc}
\end{figure}

We carried out an exhaustive search to determine the highest value of $d_\mathrm{sr}$ for which an ISR composed of $N=16$ elements will reduce the transmit power at the source to a value below that of the SISO and DF cases, for a given carrier frequency, a given rate and every $d_1\in[d_\mathrm{sr}/2,d_\mathrm{sr}]$. Fig.~\ref{fig:dsr_vs_fc} shows the highest $d_\mathrm{sr}$ values for $f_\mathrm{c}\in[2,100]$ GHz and $\bar{R}=\{5,6,7\}$ bits/sec/Hz. As expected, path loss increases with frequency and the ISR needs to be placed closer to the source. However, the slope of the curves does not change significantly over frequencies in the millimeter-wave range, i.e., $f_\mathrm{c}\in[24,100]$ GHz. Furthermore, a higher rate requires a higher transmission power, which increases the maximum value of $d_\mathrm{sr}$. To give an example, note that the maximum value of $d_\mathrm{sr}$ is $24$m for $f_\mathrm{c}=100\!$ GHz and $33$m for $f_\mathrm{c}=6\!$~GHz, when $\bar{R}=7$ bits/sec/Hz. This means that, if the IRS is placed up to $24$m away from the source, the IRS-aided system will require a lower transmit power than SISO and DF relaying for all frequencies supported by 5G. If we are interested in the sub-6 GHz range only, the IRS can be positioned up to $33$m away from the source.
 
%%%%%%%%%%%%%%%%%%%
%  Section 5. Conclusions
%%%%%%%%%%%%%%%%%%%
%\vspace{-8pt}
\section{Conclusions}
\label{sec:conclusions}

Bj\"{o}rnson~\textit{et al.}~\cite{Bjornson2020} demonstrated that intelligent reflecting surfaces cannot be seen as a panacea for wireless communication systems because they need a very large number of elements to cut down the required transmit power to a level similar to or lower than that achieved by classic single-antenna decode-and-forward relays. The results presented in this paper corroborate the findings of \cite{Bjornson2020} and highlight the differences in the performance of each scheme when the underlying channel model is changed to comply with 5G specifications.

Decode-and-forward relays exhibit clear benefits over intelligent reflecting surfaces when the direct link between a base station and a user is very weak or severely obstructed. In all other cases, we showed that intelligent reflecting surfaces composed of a reasonably low number of elements (e.g., 16) could be used to support the downlink of a base station. Therefore, decode-and-forward relays are suitable for extending the coverage area of a base station, whereas intelligent reflecting surfaces can be used to improve received signal quality or reduce transmit power requirements within the coverage area of a base station. For example, they could be placed between a base station and a relay or between successive relays in order to `guide' a signal from the base station to an intended receiver, while reducing the amount of transmitted energy that is wasted because it never reaches the receiver.

%%%%%%%%%%%%%%%%%%%
%  Acknowledgements
%%%%%%%%%%%%%%%%%%%
\vspace{3pt}
\section*{Acknowledgements}
\label{sec:ack}

The author wishes to thank Bj\"{o}rnson~\textit{et al.}~\cite{Bjornson2020} for making their simulation code publicly available.

%%%%%%%%%%%%%%%%%%%
% References
%%%%%%%%%%%%%%%%%%%
\vspace{4pt}
\bibliographystyle{IEEEtran}
\bibliography{IEEEabrv,IEEE_references}

\begin{thebibliography}{1}
\providecommand{\url}[1]{#1}
\csname url@samestyle\endcsname
\providecommand{\newblock}{\relax}
\providecommand{\bibinfo}[2]{#2}
\providecommand{\BIBentrySTDinterwordspacing}{\spaceskip=0pt\relax}
\providecommand{\BIBentryALTinterwordstretchfactor}{4}
\providecommand{\BIBentryALTinterwordspacing}{\spaceskip=\fontdimen2\font plus
\BIBentryALTinterwordstretchfactor\fontdimen3\font minus
  \fontdimen4\font\relax}
\providecommand{\BIBforeignlanguage}[2]{{%
\expandafter\ifx\csname l@#1\endcsname\relax
\typeout{** WARNING: IEEEtran.bst: No hyphenation pattern has been}%
\typeout{** loaded for the language `#1'. Using the pattern for}%
\typeout{** the default language instead.}%
\else
\language=\csname l@#1\endcsname
\fi
#2}}
\providecommand{\BIBdecl}{\relax}
\BIBdecl

\bibitem{Wu2019}
Q.~Wu and R.~Zhang, ``Intelligent reflecting surface enhanced wireless network
  via joint active and passive beamforming,'' \emph{{IEEE} Trans. Wireless
  Commun.}, vol.~18, no.~11, pp. 5394--5409, Nov. 2019.

\bibitem{Huang2019}
C.~Huang, A.~Zappone, G.~C. Alexandropoulos, M.~Debbah, and C.~Yuen,
  ``Reconfigurable intelligent surfaces for energy efficiency in wireless
  communication,'' \emph{{IEEE} Trans. Wireless Commun.}, vol.~18, no.~8, pp.
  4157--4170, Aug. 2019.

\bibitem{Liaskos2018}
C.~Liaskos, S.~Nie, A.~P. A.~Tsioliaridou, S.~Ioannidis, and I.~Akyildiz,
  ``Realizing wireless communication through software-defined hypersurface
  environments,'' in \emph{Proc. IEEE Int. Symp. World of Wireless, Mobile and
  Multimedia Networks (WoWMoM)}, Chania, Greece, Jun. 2018.

\bibitem{Ozdogan2020}
O.~\"{O}zdogan, E.~Bj\"{o}rnson, and E.~G. Larsson, ``Intelligent reflecting
  surfaces: {P}hysics, propagation, and pathloss modeling,'' \emph{{IEEE}
  Wireless Commun. Lett.}, vol.~9, no.~5, pp. 581--585, May 2020.

\bibitem{Bjornson2020}
E.~Bj\"{o}rnson, O.~\"{O}zdogan, and E.~G. Larsson, ``Intelligent reflecting
  surface versus decode-and-forward: {H}ow large surfaces are needed to beat
  relaying?'' \emph{{IEEE} Wireless Commun. Lett.}, vol.~9, no.~2, pp.
  244--248, Feb. 2020.

\bibitem{3GPP_TS_36_814}
\emph{Evolved Universal Terrestrial Radio Access ({E-UTRA}); {F}urther
  advancements for {E-UTRA} physical layer aspects}, 3GPP TS 36.814 ({V}ersion
  9.0.0, {R}elease 9), Mar. 2010.

\bibitem{3GPP_TR_38_901}
\emph{5{G}; {S}tudy on channel model for frequencies from 0.5 to 100 {GH}z},
  3GPP TR 38.901 ({V}ersion 14.3.0, {R}elease 14), Jan. 2018.

\bibitem{Rappaport2017}
T.~S. Rappaport, Y.~Xing, G.~R. MacCartney, A.~F. Molisch, E.~Mellios, and
  J.~Zhang, ``Overview of millimeter wave communications for 5{G} wireless
  networks -- with a focus on propagation models,'' \emph{{IEEE} Trans.
  Antennas Propag.}, vol.~65, no.~12, pp. 6213--6230, Dec. 2017.

\end{thebibliography}

\end{document}